\documentclass[12pt,preprint]{aastex}

\shorttitle{AL~3(BH~261): a new globular cluster in the Galaxy}
\shortauthors{Ortolani et al.}

\begin{document}


\title{AL~3 (BH~261): a new globular cluster in the Galaxy}

\author{S. Ortolani\altaffilmark{1,2,3}}

\and 

\author{E. Bica{4}}

\and

\author{B. Barbuy{5}}

\altaffiltext{1}{Visiting Astronomer, European Southern Observatory,
La Silla, Chile}
\altaffiltext{2}{The observations were carried out at the
 European Southern Observatory, La Silla, Chile, proposal 64L-0212(A)}
\altaffiltext{3}{Universit\`a di Padova, Dipartimento di Astronomia, Vicolo
 dell'Osservatorio 5, I-35122 Padova, Italy, email: ortolani@pd.astro.it}

\altaffiltext{4}{Universidade Federal do Rio Grande do Sul, 
Departamento de Astronomia, CP 15051, Porto Alegre 91501-970, Brazil,
email: bica@if.ufrgs.br}

\altaffiltext{5}{Universidade de S\~ao Paulo, Departamento de Astronomia,  
Rua do Mat\~ao 1226 S\~ao Paulo 05508-900, Brazil, 
email: barbuy@astro.iag.usp.br}

\begin{abstract}
AL~3 (BH~261), previously classified as a faint  
open cluster candidate, is shown to be a new globular cluster in the 
Milky Way, by means of B, V and I Color-Magnitude Diagrams.
The main feature of AL~3 is a prominent blue extended Horizontal Branch.
Its Color-Magnitude Diagrams
 match those of the intermediate metallicity cluster M~5. 
The cluster is projected in a rich bulge field, also contaminated by
the disk main sequence. 
The globular cluster is located  in the Galactic
 bulge at a distance from the Sun  
d$_{\odot}$ = 6.0$\pm$0.5 kpc. The reddening is E(B-V)=0.36$\pm$0.03
and the metallicity is estimated to be [Fe/H] $\approx$ -1.3$\pm$0.25. 
AL~3 is probably one of the least massive globular clusters of the Galaxy.

\end{abstract}


\section{Introduction}
New globular clusters have been discovered or identified 
in the Galaxy in the last years.
Recall  Lyng{\aa} 7 (Ortolani et al. 1993; Tavarez \& Friel 1995), 
Pyxis (Weinberger 1995; Da Costa 1995; Irwin et al. 1995; 
Sarajedini \& Geisler 1996) and IC 1257 (Harris et al. 1997).
The 2MASS survey 
has provided the two new globulars 2MASS-GC01 and 2MASS-GC02 
(Hurt et al. 2000). Ortolani et al. (2000) identified ESO 280-SC06
as a halo globular cluster projected on the disk and previously classified 
as an open cluster. Kobulnicky et al. (2005) discovered GLIMPSE-C01
from inspection of Spitzer Space Telescope images.
Carraro (2005) identified Whiting 1 (Whiting et al. 2002)
 as a young globular cluster in the halo. Finally, 
Willman et al. (2005) reported the discovery of SDSS J1049+5103,
appearing to be a globular cluster or a dwarf galaxy in the outer halo.

Harris (1996)  catalog of globular clusters
(http://www.physics.mcmaster.ca/Globular.html), as updated in February 2003,
contains 150 objects. Adding the three new objects from 2005,
 the number of galactic globular clusters amounts to 153.

AL~3 was discovered by Andrews \& Lindsay (1967) and was also cataloged as
BH~261 by van den Bergh \& Hagen (1975). It also appears in the
ESO/Uppsala catalogue (Lauberts 1982) as ESO 456-SC78.
Its angular size was estimated to be $\sim$1.5'.
The cluster was recently observed by Carraro et al. (2005) using
CCD images of 4.1'$\times$4.1', centered on the object, but they concluded
from Color-Magnitude Diagrams (CMDS)
that no cluster was present.

In the present study we show it 
to be the 154$^{\rm th}$ globular cluster in the Galaxy.
We measured the cluster center by means of 
DSS/XDSS images 
and found that it
  is located at J2000 $\alpha$ = 18$^{\rm h}$14$^{\rm m}$06.6$^{\rm s}$, 
$\delta$ = $-$28$^{\rm o}$$38'06''$, with Galactic coordinates
l = 3.36$^{\rm o}$, b = $-$5.27$^{\rm o}$. 
In Sect. 2 the observations and reductions are described. In Sect. 3
 Color Magnitude Diagrams (CMD) are presented and 
parameters are derived. 
Concluding remarks are given in Sect. 4.
 
%___
\section {Observations} 

AL~3 was observed on 2000 March 6
with the 1.54m Danish telescope  at ESO (La Silla). 
A Loral/Lesser CCD detector C1W7 with 2052$\times$2052 pixels, 
of pixel size 15 $\mu$m was used. A pixel corresponds 
to $0.39"$ on the sky, and the full 
field of the camera is $13'\times 13'$. 
Images in B (3 min.), V (1 and 3 min.) and I (10 and 40 sec.)
were obtained, under a seeing of 1.2''.
% The log of observations is provided in Table 1.
In Fig. 1 is shown a 3 min  $B$ exposure of AL~3
for a field extraction of  3.3'$\times$3.3' 
(510$\times$510 pixels).  
The image suggests  a small concentration of stars (24''$\times$12'') 
surrounded by a halo of fainter stars.
The overall appearance is similar to that of
the bulge globular cluster NGC 6540 (Bica et al. 1994), but somewhat
fainter.

Daophot II (Stetson 1987) was used to extract the instrumental magnitudes. 
For calibrations we used  Landolt (1983, 1992) standard stars.
The calibration equations are:
$V$ = 26.46 + 0.01 ($B-V$) + $v$;
$B$ = 26.40 + 0.1 ($B-V$) + $b$;
$I$ = 24.61 - 0.01 ($V-I$) + $i$,
 for 10 sec. and 15 sec. respectively, at 1.25 airmasses.
 Zero point errors are dominated by the transfer
from aperture to Daophot PSF fitting magnitudes and CCD shutter 
time uncertainties. Atmospheric extinction errors are
smaller since AL3 and  standards were observed at comparable airmasses.
Transfer errors arise from  noise in the stellar profile growth 
curve due to  crowding ($\approx$1 star per 
70 pixels). This  contaminates  stellar wings  even for the brightest and 
relatively isolated stars. We investigate   3 methods: 
(i) Statistical  reconstruction of the growth curve in small radius bins 
for the best $\approx$100 stars per 
band. (ii) Analysis of a few relatively isolated bright stars 
subtracting contaminating stars (CS) with the PSF. (iii) Growth curve analysis 
using models (Diego 1985). 
Method (i) gives the smallest errors (0.02 to 0.03 mag.), while in  
method (ii) subtraction residuals increase  noise in the star wings. This is 
caused by  possible  underestimates of the sky  around the CSs or to
PSF deviations  for faint stars and  steeper
profiles. Method (ii) should be free of statistical errors. 
It agrees  with  method (iii) within a few 
hundredths of  mag. for Lorentzian profiles. 
Gaussian profiles show $\approx$5% weaker outer wings. 
The CCD shutter time uncertainties (0.3 sec) 
related to short exposures done for standard
stars, lead to an additional 3\%
error, which is propagated to the 
calibrations of the long exposure cluster frames.
The final magnitude zero point uncertainty is estimated
to be $\pm$0.05 mag.
The atmospheric extinction was corrected with the
 La Silla standard coefficients.

\section{Color Magnitude Diagrams}

Fig. 2 shows the $V$ vs. $B-V$
CMD of  the full field. It is a typical bulge CMD similar to
that of Baade's Window (e.g. Terndrup 1988).
The prominent features are a bulge red Horizontal Branch (HB)
at V$\approx$ 16.5, $B-V$$\approx$1.5, a metal-rich Red Giant
Branch (RGB) with a turnover such as in NGC 6553 (Ortolani et al. 1990),
a disk blue main sequence, and a blue extended HB with an overdensity
at V=16, $B-V$=0.4.
In Fig. 3 we show the spatial distribution across the whole field
of the stars in a box encompassing the blue HB in the CMD. The
blue HB stars are centrally concentrated revealing  a globular cluster.
We study the stellar density profile 
derived from the BHB stars of Fig. 3, which are relatively free of 
contamination. 
The limiting radius (where the density profile mergers the background) is
r$_{\rm lim}$=3.4´$\pm$0.4´. 
The tidal radius must be larger  since 
some BHB stars reach the  field borders (Fig. 3). 
Low statistics precludes a King profile fit and deeper
images using field decontaminated TO and  MS stars 
would be necessary to explore the possibility
of post-core collapse. From the observed density  
profile we estimate a half-density radius  r$_{\rm hd}$=1´-1.5´.
 
Fig. 4 shows a V vs. $B-V$ CMD for a cluster extraction of
r$<$120 pixels (r$<$47'') and overimposed a more central one
with r$<$53 pixels (r$<$21''), of the same order of the core seen
in Fig. 1.
A few giants are present, which is typical of Palomar clusters, and the
turnoff (TO) seems to be reached near the photometry limit. 
Coordinates of the giants are available under request.
As comparison the mean locus of M5
([Fe/H]=-1.27, Harris 1996) is superimposed. This mean
locus was obtained from data taken in the same observing run,
and it is very similar to that of Sandquist et al. (1996).
The CMD of a globular cluster is clearly seen for AL~3, with a blue
Horizontal Branch, a few giants, and a populated
subgiant branch. The overall fit of M5 mean locus indicates
that AL~3 has a similar metallicity  of [Fe/H]=-1.3$\pm$0.25.

In Fig. 5 the V vs. $B-V$ offset field  at 4' north corresponding to an 
 extraction of r$<$120 pixels (r$<$47'') 
and overimposed a more central one
with r$<$53 pixels (r$<$21''), is shown.
It shows two small area samplings dominated by the bulge population.
 A comparison of Figs. 3 and 4,
shows that the evolutionary sequences are very different.
For  r$<$53 pixels (r$<$21''), we have 64 stars in the field,
and 108 in the cluster area, showing about a 4$\sigma$ excess.
In the  r$<$120 pixels (r$<$47''), we have 364 stars in the field,
and 491 in the cluster area, showing about a 7$\sigma$ excess.
We conclude that undoubtedly  we are dealing with a star cluster.
V vs. $V-I$ CMDs for cluster and field indicate essentially the
same features and conclusions.

Intermediate metallicity bulge globular clusters that are relatively
depleted in giants, with blue extended HBs are in general
related to post-core collapse structure (Fusi Pecci et al. 1993; 
Trager et al. 1995).
Examples are NGC 6522 (Barbuy et al. 1994; Terndrup et al. 1998),
 NGC 6540 (Bica et al. 1994), HP 1 (Barbuy et al. 2006; 
Ortolani et al. 1997) and NGC 6558 (Rich et al. 1998).
AL~3 appears to be one more such cluster.

Fig. 6 shows I vs. $B-I$ for the same extractions as in Fig. 4,
where again the M5 mean locus is superimposed.
The blue stragglers stars (BSS) sequence of M5 is also shown, indicating
that some BSS are present in AL~3.
The good fit of the BSS locus between AL~3 and M5 suggests a comparable
age for the clusters. Despite not having a clear
turn-off (TO) locus, the locus of BSS is a prolongation of
the main sequence, providing information on the TO location.

\subsection{Reddening and distance for AL~3}

The fit of M5 sequences to those of AL~3 in Fig. 4  
gives
 V$_{\rm HB}$ = 15.7$\pm$0.2 for the brightest part of
the BHB. The color of the RGB at the HB level is 
$B-V$ = 1.16$\pm$0.05. The difference with respect
to the fit with M5 is 
 $\Delta$(B-V)$_{\rm (AL~3 - M5)}$ = 0.33,
so that the reddening of AL~3 is E(B-V)=0.36
considering   E(B$-$V) = 0.03 for M5 (Harris 1996).
Adopting R = 3.1 we obtain A$_{\rm V}$ = 1.12, and using
M$_{\rm V}^{\rm HB}$ = 0.7 suitable for the cluster metallicity 
(Buonanno et al. 1989), there results (m-M)$_{\rm o}$ = 13.9,   
and the distance to the Sun d$_{\odot}$ = 6.0$\pm$0.5 kpc 
for AL~3.
  Assuming the distance of the Sun to  the Galaxy center to be 
R$_{\odot}$ = 8.0 kpc (Reid 1993), the Galactocentric coordinates 
are  X = -2.0 kpc (X $>$ 0 is on the other side of the Galaxy),
Y = $-$0.35 kpc and Z = $-$0.55 kpc. The distance from the Galactic 
center is R$_{\rm GC}$ = 2.1 kpc, thus AL~3 is
an inner bulge cluster, and should be incorporated to the 
sample of Barbuy et al. (1998), both in terms of 
angular and radius separation from the center.
Bica et al. (2006) derived a distance of the Sun to  the Galaxy center
of R$_{\odot}$ = 7.2 kpc, using globular clusters. In this case,
X = -2.0 kpc, Y = $-$0.35 kpc and Z = $-$0.55 kpc, and
 R$_{\rm GC}$ = 1.2 kpc, and
 AL~3 is even closer to the center.

We estimated the cluster observed integrated magnitude in the V 60 sec.  
image, by flux integration  within r $<$120"  and subtraction of  the 
background measured around the cluster in a factor 6  larger area. 
The sky background error is $\approx$5\%, while the average cluster flux 
is only 10\% above the background.  We derived V$_{\rm t}$=11.0 $\pm$0.4
and with  reddening and distance values as above we obtained  
M$_{\rm V}$=-4.0$\pm$0.45  for AL 3. Comparing to Harris´s (1996) 
compilation we conclude that AL 3 is among the intrisically faintest
globular clusters, only somewhat more luminous than  AM 4, Pal 1, E3 
and Pal 13.

\section{Concluding remarks}

We identified the object AL~3 discovered by Andrews \& Lindsay (1967)
as  a new globular cluster in the Galaxy. 
Since the discovery of Djorgovski 1, and identification
of ESO 456-SC38 and NGC 6540  
(Djorgovski 1987), AL~3 is the next inner bulge
globular cluster identified.
 We find a metallicity of [Fe/H] $\approx$ -1.3$\pm$0.25.  
 It is  located in the inner bulge, at a distance d$_{\odot}$ = 6$\pm$0.5 kpc.
It appears to be another bulge intermediate metallicity globular
 cluster, with a blue extended Horizontal Branch,
and a depleted red giant branch (Rich et al. 1998).
 AL~3 Color-Magnitude
Diagrams suggest that the cluster may have about 10 giants,
similar to Palomar 13 (Siegel et al. 2001), and somewhat more massive
than AM~4 which has no giants and a mass less than 1000 M$_{\odot}$
(Inman \& Carney 1987).
 We are dealing with one of the least massive globular
clusters, similar to several Palomar clusters.
 The contrast
in metallicity between AL~3 and the bulge  made possible the
identification of the cluster, a metal
rich globular cluster in the same conditions would  be indistinguishable
from the bulge population.    
Other faint globular clusters may still be found in the bulge.
 AL~3, as a new intermediate metallicity globular cluster located in
the bulge, is another interesting target for abundance determinations
and chemical enrichment studies of the Galaxy.

\begin{acknowledgements}

We acknowledge partial financial support from
the brazilian agencies CNPq and Fapesp,
and Ministero dell'Universit\`a e della Ricerca
Scientifica e Tecnologica (MURST) under the program
on 'Fasi iniziali di Evoluzione dell'Alone e del Bulge Galattico' (Italy).
\end{acknowledgements}

%--------------------------------- References -------------

%

\clearpage

\begin{figure}
\includegraphics[angle=0,scale=.50]{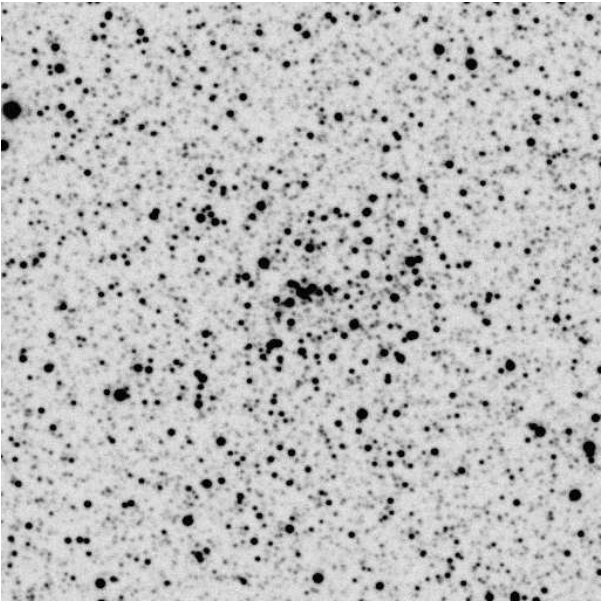}
\caption{AL~3: 3 min. B image. Extraction of 3'$\times$3'.
North is up and East is left.}
\end{figure}

\clearpage

\begin{figure}
\includegraphics[angle=0,scale=.50]{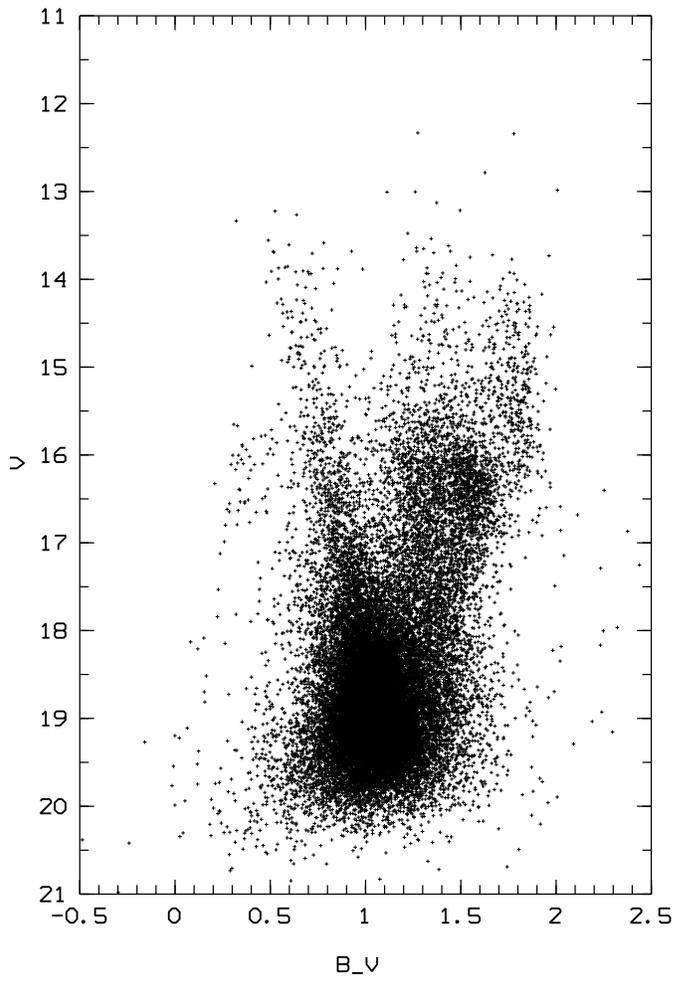}
\caption{V vs. $B-V$ for whole field (13'$\times$13').}
\end{figure}

\clearpage

\begin{figure}
\includegraphics[angle=0,scale=.50]{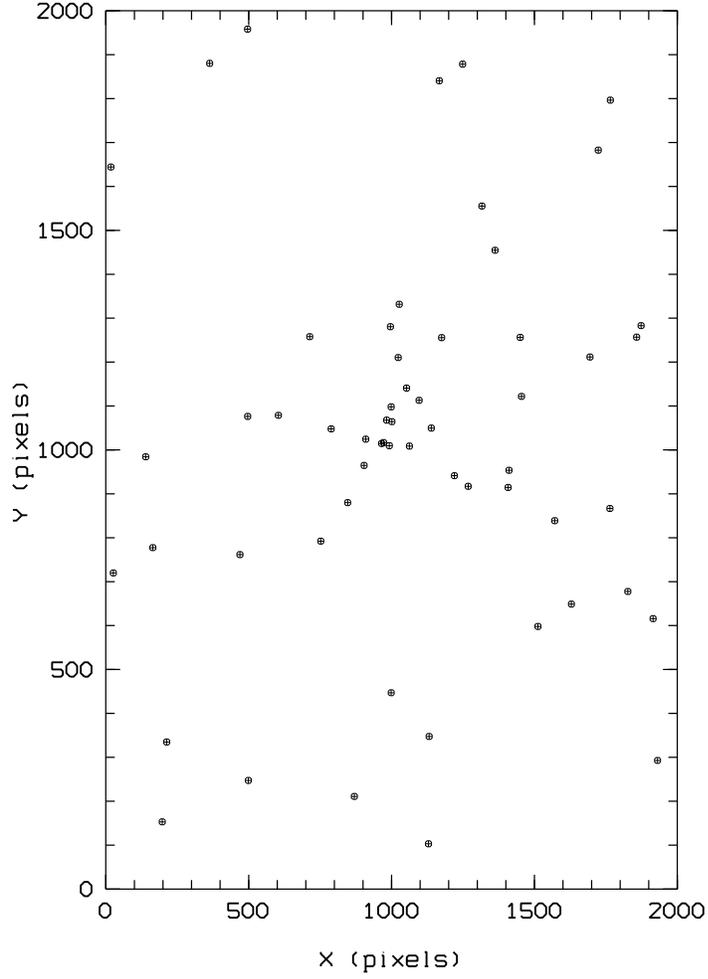}
\caption{Whole field (13'$\times$13') in X, Y pixels
showing the distribution of blue Horizontal Branch stars,
defined in the box 15.5$<$V$<$17.1 and 0$<$B-V$<$0.55.}
\end{figure}

\clearpage

\begin{figure}
\includegraphics[angle=0,scale=.50]{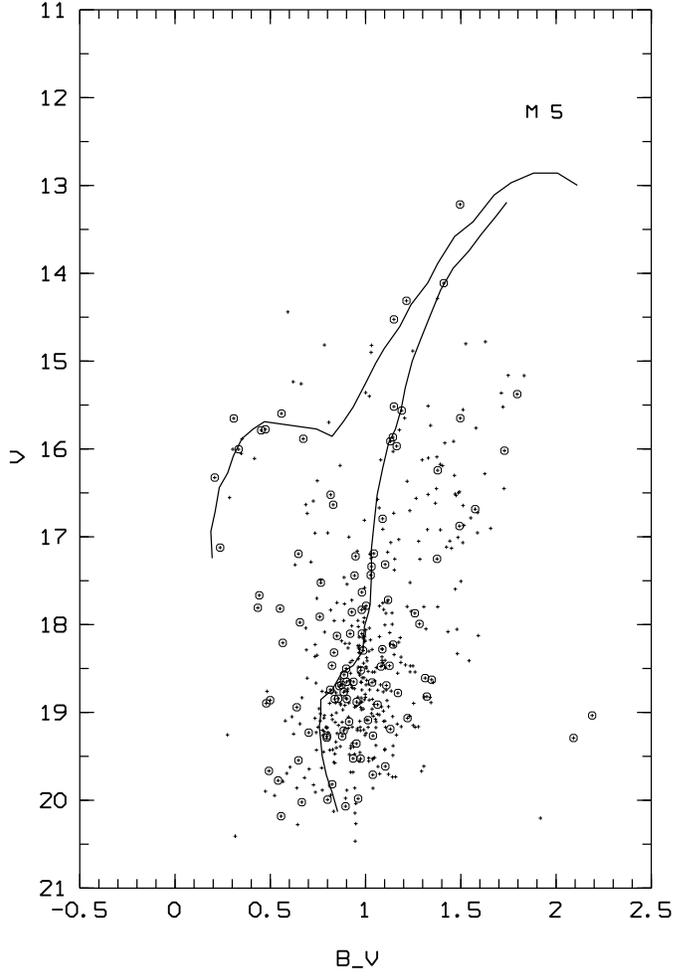}
\caption{AL~3: V vs. $B-V$ for extractions of 
r$<$120 pixels (r$<$47''): crosses;
 and overimposed 
r$<$53 pixels (r$<$21''): open circles.
M5 mean locus is superimposed.
}
\end{figure}

\clearpage

\begin{figure}
\includegraphics[angle=0,scale=.50]{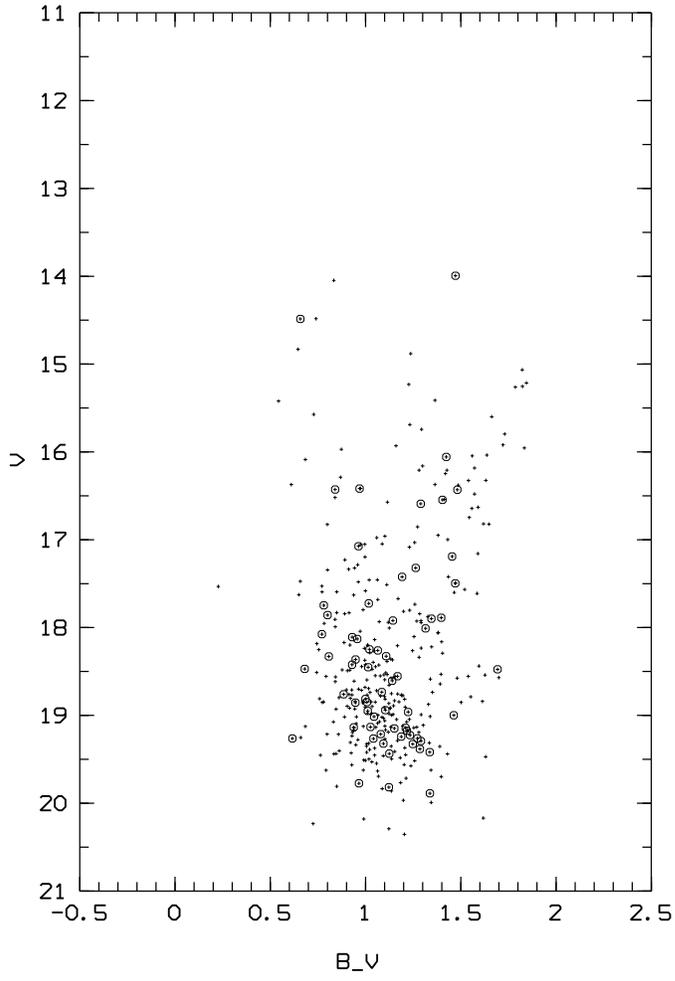}
\caption{Same as Fig. 4 for offset field 4' north.
}
\end{figure}

\clearpage

\begin{figure}
\includegraphics[angle=0,scale=.50]{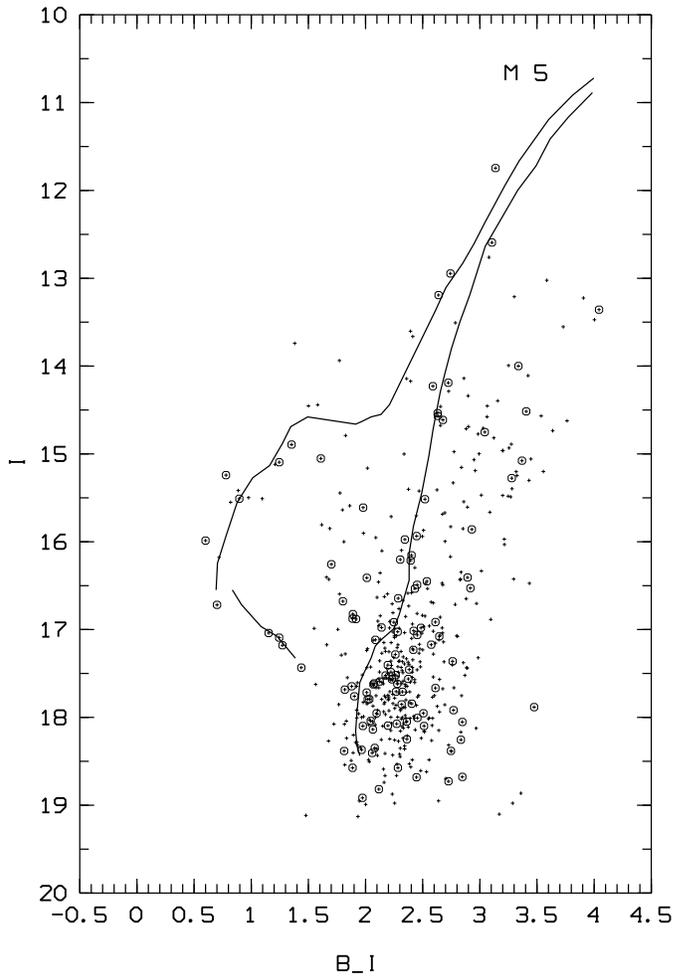}
\caption{AL~3: Same as Fig. 4 in I vs. $B-I$.
}
\end{figure}

\end{document}